\documentclass[11pt,a4paper]{article}
\usepackage{jheppub}

\usepackage{amsmath}
\usepackage{latexsym}

\title{\Large{Noncommutative cosmological models coupled to a perfect fluid and a cosmological constant}}

\author[a,b,c]{E. M. C. Abreu,}
\author[c]{M. V. Marcial,}
\author[c]{A. C. R. Mendes,}
\author[c]{W. Oliveira}
\author[c]{and G. Oliveira-Neto}

\affiliation[a]{Grupo de F\'isica Te\'orica, Departamento de F\'{\i}sica, \\
Universidade Federal Rural do Rio de Janeiro\\
BR 465 km 07, 23890-971, Serop\'edica, RJ, Brazil}
\affiliation[b]{Centro Brasileiro de Pesquisas F\'isicas (CBPF),\\
Rua Xavier Sigaud, 150, 22290-180, Rio de Janeiro, Brazil}
\affiliation[c]{Departamento de F\'{\i}sica, ICE, Universidade Federal de Juiz de Fora,\\
36036-330, Juiz de Fora, MG, Brazil}
\emailAdd{evertonabreu@ufrrj.br}
\emailAdd{mateusmarcial@fisica.ufjf.br}
\emailAdd{albert@fisica.ufjf.br}
\emailAdd{wilson@fisica.ufjf.br}
\emailAdd{gilneto@fisica.ufjf.br}
\abstract{In this work we carry out a noncommutative analysis of several
Friedmann-Robert-Walker models, coupled to different types of perfect fluids
and in the presence of a cosmological constant. The classical field equations
are modified, by the introduction of a shift operator, in order to introduce
noncommutativity in these models. We notice that the noncommutative versions of
these models show several relevant differences with respect to the correspondent
commutative ones.}


\usepackage{graphicx}% Include figure files
\usepackage{dcolumn}% Align table columns on decimal point
\usepackage{bm}% bold math
\usepackage{amsmath}
\usepackage{latexsym}
\newcommand{\be}{\begin{equation}}
\newcommand{\ee}{\end{equation}}
\newcommand{\ea}{\end{eqnarray}}
\newcommand{\ba}{\begin{eqnarray}}
\def\ni{\noindent}

\def\[{\left\lbrack}
\def\]{\right\rbrack}

\def\({\left(}
\def\){\right)}

\begin{document}

\maketitle
\flushbottom

\pagestyle{myheadings}
\markright{Noncommutative cosmological models coupled...}

%\newpage

%\setlength{\baselineskip} {20 pt}

\section{Introduction}

In current theoretical physics there is a relevant number of theoretical
investigations that lead us to believe that at the Big-Bang first moments, 
the geometry was not commutative and the dominating physics at
that time was ruled by the laws of noncommutative (NC) geometry. Therefore,
the idea is that the physics of the early moments can be constructed based on
these concepts.

The first published steps through this knowledge were given by Snyder
\cite{snyder} who believes that NC principles could make the quantum
field theory infinities disappear.  However, it was not accomplished
\cite{yang} and Snyder's ideas were put to sleep for a long time.
The main modern motivations that rekindle the investigation about NC field
theories came from string theory and quantum gravity \cite{strings}.

In the context of quantum mechanics for example, R. Banerjee \cite{banerjee}
discussed how NC structures appear in planar quantum mechanics
providing a useful way for obtaining them. The analysis was based on the
NC algebra in planar quantum mechanics that was originated
from 't Hooft's analysis on dissipation and quantization \cite{thooft}.

It is opportune to mention here that this noncommutativity in the context
of string theory mentioned above could be eliminated constructing a
mechanical system which reproduces the string classical dynamics \cite{string}.
NC field theories have been studied intensively in many branches of physics
\cite{witten}-\cite{hp}.
%,other,alexei,belov,szabo,omer,RB,hp}.

In a very interesting paper, a parallel investigation was developed by Duval
and Horv\'{a}thy \cite{dh}, where it was obtained the anomalous commutation
relations for the coordinates obtained through the ``Peierls substitution"
\cite{peierls}. From first principles, without using such unphysical limit,
the authors introduced NC (quantum) mechanics starting with group theory and
applied it to condensed matter physics, e.g., the Hall effect. The respective
Lagrangian approach was discussed in detail in subsequent papers \cite{dh2}.
Dunne, Jackiw and Trugenberger \cite{djt} justify the Peierls rule by considering
the $m\rightarrow 0$ limit, reducing the classical phase space from four to two
dimensions, parameterized by NC coordinates $X$ and $Y$, whereas the potential
becomes an effective Hamiltonian.

In \cite{mrs} the authors analyzed the IR and UV divergences and verified that Planck's
constant enters via loop expansion. Here, differently, we make a
non-perturbative approach and we will see that Planck's constant enters
naturally in the theory via Moyal-Weyl product.

A general algebra $\alpha$-deformation of classical
observables that introduces a general NC quantum mechanics was constructed in \cite{DJEMAI1}.
This $\alpha$-deformation is equivalent to some general transformation for
the usual quantum phase space variables.  In other words, the authors discuss
the passage from classical mechanics to quantum mechanics. Then to NC quantum
mechanics, which allows to obtain the associated NC classical mechanics. This
is possible since quantum mechanics is naturally interpreted as a NC (matrix)
symplectic geometry \cite{DJEMAI2}.

In \cite{amorim}, the author constructed an extension of the well known
Doplicher-Fredenhagen-Roberts NC algebra introducing the formalism which is
now called in the literature as the Doplicher-Fredenhagen-Roberts-Amorim algebra.
In this formalism the NC parameter $(\theta)$ is an ordinary coordinate of the
spacetime and therefore it has a canonical conjugate momentum $(\pi)$. An
extended Hilbert space was constructed together with all the ingredients of a
new NC quantum mechanics. 
But notice that both preserves the underlying NC relation
$[x^\mu ,x^\nu ]=i\theta^{\mu\nu}$.  For details, the interested reader can
consult \cite{sigma}.

Back to our main subject here, in few words we can say that one way to obtain NC
versions for field theories one have to replace the usual product of fields
into the action by the Moyal-Weyl product, defined as
\begin{equation}
\label{primeira}
\phi_1 (x)\star \phi_2 (x) \,=\,exp\left( {i\over 2} \theta^{\mu \nu}
\partial^{x}_{\mu}\partial^{y}_{\nu} \right) \phi_1 (x) \phi_2 (y)\mid_{x=y},
\end{equation}
where $\theta^{\mu \nu}$ is a real and antisymmetric constant matrix. As a
consequence, NC theories are highly nonlocal. We also note a basic NC property
that the Moyal-Weyl product of two fields inside the action is the same as the
usual product, considering that we discard boundary terms. Thus, the
noncommutativity affects just the vertices.

Some years back, \cite{sympnc} three of us have proposed a new formalism to generalize the
quantization by deformation introduced in \cite{DJEMAI1} in order to explore,
with a new insight, how the NC geometry can be introduced into a (commutative)
field theory. To accomplish this, a systematic way to introduce NC geometry into
commutative systems, based on Faddeev-Jackiw symplectic formalism and Moyal-Weyl
product, was presented \cite{amo}.

One important arena where NC ideas may play an important role
is cosmology. If superstrings is the correct theory to unify all the interactions
in nature, it must have played the dominant role at very early stages of our Universe.
At that time, all the canonical variables and corresponding momenta describing
our Universe should have followed a NC algebra. Inspired by these
ideas some researchers have considered such NC models in quantum
cosmology \cite{garcia,nelson,barbosa}. It is also possible that
some residual NC contribution may have survived in later stages of
our Universe. Based on these ideas some researchers have proposed some NC
models in classical cosmology in order to explain some intriguing results observed by
WMAP. Such as a running spectral index of the scalar fluctuations and an anomalously
low quadrupole of CMB angular power spectrum. Among such proposals we may mention the
following ones \cite{huang,kim,liu,huang1,kim1}. Another
relevant application of the NC ideas in classical cosmology is the attempt
to explain the present accelerated expansion of our Universe \cite{pedram,obregon,neves}.

We have organized this paper as follows. In section \ref{section2}, we introduce the
generalized quantization by deformation formalism assuming a generic classical symplectic
structure. We will construct a new star product which appears at first sight to suffer from the
non-associative property disease.  We will demonstrate that we can recover
this crucial property for the star product. In section \ref{section3}, we construct the
NC versions of several Friedmann-Robertson-Walker cosmological models.
These models may have positive, negative or zero spatial sections curvatures, they
are coupled to different types of perfect fluids and they may have a positive, negative
or zero cosmological constant. We have used the Schutz variational formalism in order to
write the Hamiltonian of the models. In section \ref{section4}, we perform a complete
study of the evolution of the universes described by each model. We solve the dynamical
equations for the scale factor. Based on our results, we conclude how the NC
parameter modifies the evolution of the universe in different models. Also in this
section, we present some special cases where our conclusions can be clearly verified.
In section \ref{section5}, based on our previous results, we write a function ($\widetilde{\Lambda}$)
that depends on three parameters that generalizes the cosmological constant.  We discuss
the possible scenarios for the Universe evolution predicted by $\widetilde{\Lambda}$, depending
on the value of $\alpha$. The final considerations and the conclusions are depicted in the last section.

\section{The NC Generalized Symplectic Formalism}
\label{section2}

The quantization by deformation \cite{MOYAL1} consists in the substitution of
the canonical quantization process by the algebra ${\cal A}_\hbar$ of quantum
observables generated by the same classical one obeying Moyal-Weyl product,
{\it i.e.}, the canonical quantization
\begin{equation}
\label{II1}
\lbrace h, g\rbrace_{PB} = \frac{\partial h}{\partial \zeta_a} \omega_{ab}
\frac{\partial g}{\partial \zeta_b}  \longrightarrow \frac {1}{\imath\hbar}
[{\cal O}_h, {\cal O}_g]\;\;,
\end{equation}

\noindent with $\zeta=(q_i,p_i)$, is replaced by the $\hbar$-star deformation
of ${\cal A}_0$, given by

\begin{equation}
\label{II2}
\lbrace h, g\rbrace_{\hbar} = h\star_\hbar g - g \star_\hbar h\;\;,
\end{equation}
where
\begin{equation}
\label{II3}
(h \star_\hbar g)(\zeta)\,=\,\exp\{\frac{\imath}{2}\hbar \omega_{ab}\partial^a_{(\zeta_1)}\partial^b_{(\zeta_2)}\}h(\zeta_1)g(\zeta_2)|_{\zeta_1=\zeta_2=\zeta}
\;\;,
\end{equation}
with $a,b=1,2,\dots,2N$ and with the following classical symplectic structure
\begin{equation}
\label{II4}
\omega_{ab} = \left( \begin{array}{cc}
0 & \delta_{ij} \\
-\delta_{ji} & 0
\end{array} \right)
\,\,\,{\text with}\,\,\, i,j=1,2,\dots,N\;\;,
\end{equation}
which satisfies the relation
\begin{equation}
\label{II5}
\omega^{ab}\omega_{bc} = \delta^a_c\;\;.
\end{equation}

In the next section we will describe the conceptual basis that support our
method and our cherished results.  However, we think that it is crucial for
the reader to understand mathematically the underlying equations used in this
work.  We will show that the crucial property of associativity is not lost in
our formalism.

\subsection{Deformation quantization}

Notice that we have a subtle conceptual analogy between both star products,
i.e., the one defined by the Moyal-Weyl product in Eq. (\ref{primeira}) and
the $\Sigma$-star deformation product defined in Eq. (\ref{II3}). In general
the star products so defined are associative if the parameter is constant.
If we do not have associativity, the product so defined is useless in physics.
We will talk with more detail about this from now on.

Kontsevich proved in \cite{Kontsevich} that any finite-dimensional Poisson
manifold can be canonically quantized (deformation quantization). We will write
now a quite brief introduction to deformation quantization
following \cite{Kontsevich}.

Let us define an algebra $A=\Gamma(X,{\cal O}_X)$ over ${\cal R}$ of smooth
functions on a finite-dimensions $C^{\infty}$-manifold $X$.  We construct a
star product on $A$ defined as being an associative $R[\hbar]$-linear product
on $A[\hbar]$. This star product between $f$ and $g$
($f,g\,\in\,A\,\subset\,A[\hbar]$) can be written as

\begin{equation}
\label{AAA}
(f,g) \rightarrow f \star g\,=\, fg\,+\,\hbar\,B_1\,(f,g)\,+\,\hbar^2\,B_2\,(f,g)\,
+\ldots \in A[\hbar]
\end{equation}
where $\hbar$ is a constant parameter and $B_i$ are bidifferential operators.
A bidifferential operator can be understood as bilinear maps which are
differential operator \cite{Kontsevich}. And the product of arbitrary elements
of $A[\hbar]$ can be defined following the framework written in (\ref{AAA}) and
it is defined by the condition of linearity over ${\cal R}[\hbar]$

\begin{equation}
\label{BBB}
\Big(\,\sum_{n\geq 0}\,f_n\,\hbar^n \Big) \star \Big(\,\sum_{n \geq 0}\,g_n\,\hbar^n \Big)
\,=\,\sum_{k,l\geq 0}\,f_k\,g_l\,\hbar^{k+l}\,+\,\sum_{k,l \geq 0, m\geq l}\,B_M\,(f_k,g_l)\,\hbar^{k+l+m}\,\,.
\nonumber \\
\end{equation}
For more details the interested reader can look in \cite{Kontsevich}.

The simplest of a deformation quantization is the Moyal-Weyl product for the
Poisson structure on ${\cal R}^d$ with constant coefficients,
%\begin{widetext}
\begin{eqnarray}
\label{CCC}
&&f\star g\,+\,f\,g\,+\,\hbar\,\sum_{i,j}\,\alpha^{ij}\,\partial_i (f)\,\partial_j (g)
\,+\,\frac{\hbar^2}{2}\sum_{i,j,k,l}\,\alpha^{ij}\,\alpha^{kl}\,\partial_i
\partial_k (f)\,\partial_j \partial_l (g)\,+\, \ldots \nonumber \\
&=&\sum^{\infty}_{n=0}\,\frac{\hbar^n}{n!}\sum_{i_1,\ldots,i_n,j_1,\ldots,j_n}\,\prod^n_{k=1}\,\alpha^{i_k\,j_k}
\Big(\prod^n_{k=1}\partial_{i_k} \Big)(f) \cdot \Big(\prod^n_{k=1}\partial_{j_k} \Big)(g)
\end{eqnarray}
where $\alpha^{ij}\,=\,-\alpha^{ji}$.
%\end{widetext}

Let $\alpha\,=\,\sum_{i,j}\,\alpha^{ij}\,\partial_i \wedge \partial_J$ be a
Poisson bracket with variable coefficients in an open domain of ${\cal R}^d$
\cite{Kontsevich}.  Namely, $\alpha^{ij}$ is not a constant but a function of
coordinates. Then the following star product gives an associative product
modulo $O(\hbar^3$) \cite{Kontsevich},

\begin{eqnarray}
\label{DDD}
&&f \star g\,=\,f\,g\,+\,\hbar\,\sum_{i,j}\,\alpha^{ij}\,\partial_i (f)\,\partial_j (g)
\,+\,\frac{\hbar^2}{2}\sum{i,j,k,l}\,\alpha^{ij}\alpha^{kl}\,\partial_i\,\partial_k (f)\,
\partial_j \partial_l (g) \nonumber \\
&+&\frac{\hbar^3}{3}\,\sum_{i,j,k,l}\,\alpha^{ij}\,\partial_j (\alpha^{kl}) \Big(\partial_i\,
\partial_k (f)\,\partial_l (g)\,-\, \partial_k (f) \partial_i \partial_l (g) \Big)
\,+\, O(\hbar^3) \,\,.
\end{eqnarray}
For us to demonstrate the associativity up to the second order means that for
any three functions $f,g$ and $h$ we have that

\begin{equation}
\label{EEE}
(f \star g) \star h \,=\, f \star (g \star h)
\end{equation}

We can see clearly form (\ref{DDD}) that the fact of $\alpha^{ij}$ being a
non-constant parameter brings a different $\hbar^3$ order for the Moyal-Weyl
product. A simple comparison between Eqs. (\ref{CCC}) and (\ref{DDD}) can
make us see that if we can not write a star product like the one described
in (\ref{DDD}) so we do not have associativity. We will turn back to this
issue in a few moments.

The quantization by deformation can be generalized assuming a generic
classical symplectic structure $\Sigma^{ab}$. In this way the internal
law will be characterized by $\hbar$ and by another deformation parameter
(or more). As a consequence, the $\Sigma$-star deformation of the algebra
becomes
\begin{equation}
\label{II6}
(h \star_{\hbar\Sigma} g)(\zeta)\,=\, \exp\{\frac{\imath}{2}\hbar \Sigma_{ab}\partial^a_{(\zeta_1)}\partial^b_{(\zeta_2)}\}h(\zeta_1)g(\zeta_2)|_{\zeta_1=\zeta_2=\zeta}\,\,,
\end{equation}
with $a,b=1,2,\dots,2N$.

This new star-product generalizes the algebra among the symplectic
variables in the following way
\begin{equation}
\label{II7}
\lbrace h, g\rbrace_{\hbar\Sigma} = \imath\hbar\Sigma_{ab}\;\;.
\end{equation}

Notice that the new star product in (\ref{II6}) is defined in an analogous
way as the Moyal-Weyl product. However, in Eq. (\ref{II7}) we see that the
parameter $\Sigma_{ab}$ is not constant. Hence, we can realize that
from Eqs. (\ref{II6}) and (\ref{II7}) the associativity property was lost
in Eq. (\ref{II6}).

However, if we consider only terms up to $\hbar^2$ and due to the smallness
of $\hbar$ we can recover the associativity property of Eq. (\ref{II6}).
So, the general expression given in Eq. (\ref{II6}) hides the fact that the
expansion of the exponential is physically valid only for terms proportional
to $1, \hbar$ and $\hbar^2$ and therefore it is not completely correct. Our
objective in Eq. (\ref{II6}) was to write the $\Sigma$-star product in a
compact way.  But we have to make this observation.
%Notice that in Eqs. (\ref{AAA})-(\ref{DDD}), $\hbar$ was just a constant.  And in (\ref{II6}) it is the Planck constant.

On the other hand, we will see soon that Eq. (\ref{II6}) is not the starting
point of our procedure. The main point is the generalized Dirac quantization
defined in the next section.

The motivation to construct a $\Sigma$-star deformed product like the one in
Eq. (\ref{II6}) is to show that we can have in physics an alternative and at
the same time associative star product with variable coefficients without
losing the associativity property as we showed above, i.e., where the expansion
stops in the second order of $\hbar$.

\subsection{The NC approach}

In \cite{DJEMAI1,DJEMAI2}, the authors proposed a quantization process to
transform the NC classical mechanics into the NC quantum mechanics, through
generalized Dirac quantization,

\begin{equation}
\label{II8}
\lbrace h, g\rbrace_{\Sigma} = \frac{\partial h}{\partial \zeta_a} \Sigma_{ab}
\frac{\partial g}{\partial \zeta_b}  \longrightarrow \frac {1}{\imath\hbar}
[{\cal O}_h, {\cal O}_g]_{\Sigma}\;\;.
\end{equation}

\noindent The relation above can also be obtained through a particular
transformation onto the usual classical phase space, namely,
\begin{equation}
\label{II9}
\zeta^\prime_a = T_{ab} \zeta^b\;\;,
\end{equation}
where the transformation matrix is
\begin{equation}
\label{II10}
T = \left( \begin{array}{cc}
 \delta_{ij} & - \frac{1}{12} \theta_{ij} \\
\frac{1}{12} \beta_{ij}  & \delta_{ij}
\end{array} \right)\;\;,
\end{equation}
with $\theta_{ij}$ and $\beta_{ij}$ being antisymmetric matrices. As a
consequence, the original Hamiltonian becomes
\begin{equation}
\label{II11}
{\cal H}(\zeta_a) \longrightarrow {\cal H}(\zeta^\prime_a)\;\;.
\end{equation}
The corresponding symplectic structure is
\begin{equation}
\label{II12}
\Sigma_{ab} = \left( \begin{array}{cc}
\theta_{ij} & \delta_{ij}+\sigma_{ij} \\
-\delta_{ij}-\sigma_{ij} & \beta_{ij}
\end{array} \right) \;\;,
\end{equation}
$\sigma_{ij} = - \frac{1}{18} [\theta_{ik}\beta_{kj} + \beta_{ik}\theta_{kj}]$.
Due to this, the commutator relations look like
\begin{eqnarray}
\label{II13}
\[q^\prime_i, q^\prime_j\] &=& \imath\hbar\theta_{ij}\;\;,\nonumber\\
\[q^\prime_i, p^\prime_j\] &=& \imath\hbar (\delta_{ij} + \sigma_{ij})\;\;,\\
\[p^\prime_i, p^\prime_j\] &=& \imath\hbar\beta_{ij}\;\;.\nonumber
\end{eqnarray}

At this point, it is important to notice that a Lagrangian formulation
was not given so far. Now, we propose a new systematic way to obtain a
NC Lagrangian description for a commutative system. In order to achieve
our objective, the symplectic structure $\Sigma_{ab}$ must firstly be
fixed and after that, the inverse of $\Sigma_{ab}$ must be computed.
As a consequence, an interesting problem arise: if there are some
constants ({\it Casimir invariants}) in the system, the symplectic
structure has a zero-mode, given by the gradient of these {\it Casimir
invariants}. Hence, it is not possible to compute the inverse of
$\Sigma_{ab}$. However, in Ref. \cite{CNWO} this kind of problem was
solved. On the other hand, if $\Sigma_{ab}$ is nonsingular, its inverse
can be obtained solving the next relation
\begin{equation}
\label{II14}
\int{\Sigma_{ab}(x,y) \, \Sigma^{bc}(y,z) dy}\, = \,\delta_a^c \delta(x-z)\;\;,
\end{equation}

\noindent which generates a set of differential equations since $\Sigma^{ab}$
is an unknown two-form symplectic tensor obtained from the following first-order
Lagrangian
\be
\label{II15}
{\cal L} = A_{\zeta^\prime_a} \dot\zeta^{\prime a} - V(\zeta^\prime_a)\;\;,
\ee
as being
\be
\label{II16}
\Sigma^{ab}(x,y) = \frac {\delta A_{\zeta^\prime_a}(x)}{\delta \zeta^\prime_b(y)}
- \frac {\delta A_{\zeta^\prime_b}(x)}{\delta \zeta^\prime_a(y)}\;\;.
\ee

\noindent Due to this, the one-form symplectic tensor, $A_{\zeta^\prime_a}(x)$,
can be computed and subsequently, the Lagrangian description, Eq. (\ref{II15}),
is obtained also.

In order to compute $A_{\zeta^\prime_a}(x)$, the Eq. (\ref{II14}) and
Eq. (\ref{II16}) will be used, which generates the following set of differential
equations
\ba
\label{II17}
\theta_{ij} B_{jk}(x,y) + \(\delta_{ij}+\sigma_{ij}\)A_{jk}(x,y) &=& \delta_{ik}\delta(x-y)\;\;,\nonumber\\
A_{jk}(x,y) \theta_{ji} + \(\delta_{ij}+\sigma_{ij}\)C_{jk}(x,y) &=& 0\;\;,\nonumber\\
- \(\delta_{ij} + \sigma_{ij}\)B_{jk}(x,y) + \beta_{ij}A_{jk}(x,y) &=& 0\;\;, \\
A_{kj}(x,y)\(\delta_{ji} + \sigma_{ji}\) + \beta_{ij} C_{jk}(x,y) &=& \delta_{ik}\delta(x-y)\;\;,\nonumber
\ea
where
\ba
\label{II18}
B_{jk}(x,y) &=& \(\frac {\delta A_{q^\prime_j}(x)}{\delta q^\prime_k(y)} - \frac {\delta A_{q^\prime_k}(x)}{\delta q^\prime_j(y)}\)\;\;,\nonumber\\
A_{jk}(x,y) &=& \(\frac {\delta A_{p^\prime_j}(x)}{\delta q^\prime_k(y)} - \frac {\delta A_{q^\prime_k}(x)}{\delta p^\prime_j(y)}\)\;\;,\nonumber\\
C_{jk}(x,y) &=& \(\frac {\delta A_{p^\prime_j}(x)}{\delta p^\prime_k(y)} - \frac {\delta A_{p^\prime_k}(x)}{\delta p^\prime_j(y)}\)\;\;.
\ea

\noindent From the set of differential equations in Eq. (\ref{II17}),
and the equations above, Eq. (\ref{II18}), we compute the quantities
$A_{\zeta^\prime_a}(x)$.

As a consequence, the first-order Lagrangian can be written as

\be
\label{II19}
{\cal L} = A_{\zeta^\prime_a} \dot\zeta^\prime_a - V(\zeta^\prime_a)\;\;.
\ee
Notice that, despite (\ref{II15}) and (\ref{II19}) have the same form,
in (\ref{II19}) the $A_{{\zeta'}_a}$ are completely computed through
the solution of the system (\ref{II17}). In both we have a NC version
of the theory as a consequence of the deformation in (\ref{II10}) and
its corresponding symplectic structure in (\ref{II12}).

\section{The noncommutative Friedmann-Robertson-Walker models}
\label{section3}

Friedmann-Robertson-Walker (FRW) cosmological models are characterized by the
scale factor $a(t)$ and have the following line element,

\begin{equation}
\label{1}
ds^2 = - N(t)^2 dt^2 + a(t)^2\left( \frac{dr^2}{1 - kr^2} + r^2 d\Omega^2
\right)\, ,
\end{equation}
where $d\Omega^2$ is the line element of the two-dimensional sphere with
unitary radius, $N(t)$ is the lapse function and $k$ gives the type of
constant curvature of the spatial sections. The curvature is positive for $%
k=1$, negative for $k=-1$ and zero for $k=0$. Here, we are using the natural
unit system, where $c = 8\pi G = 1$.  We assume that the matter content of the model is
represented by a perfect fluid with four-velocity $U^\mu = \delta^{\mu}_0$
in the co-moving coordinate system used, plus a cosmological constant ($\Lambda$)
which can be either positive, negative or zero. The total energy-momentum
tensor is given by,

\begin{equation}
T_{\mu\nu} = (\rho+p)U_{\mu}U_{\nu} - p g_{\mu\nu} - \Lambda
g_{\mu\nu}\, ,
\label{2}
\end{equation}
where $\rho$ and $p$ are the energy density and pressure of the fluid,
respectively. Here, we assume that $p = \alpha\rho$, which is the equation of
state for a perfect fluid.

Einstein's equations for the metric (\ref{1}) and the energy momentum tensor
(\ref{2}) are equivalent to the Hamilton equations generated by the
following super-Hamiltonian constraint \cite{germano1},

\begin{equation}
\label{3}
{\mathcal{H}}= -\frac{P_{a}^2}{12a} - 3ka +\Lambda a^{3} + P_{T}a^{-3\alpha}\,\,,
\end{equation}
where $P_{a}$ and $P_{T}$ are the momenta canonically conjugated to $a$ and $
T$ respectively, the latter being the canonical variable associated to the fluid. This
super-Hamiltonian was derived using the Schutz variational formalism \cite{schutz,schutz1}.
%The classical dynamics is governed by the Hamilton equations, derived from
%eq. (\ref{3}) and the constraint equation $\mathcal{H} = 0$. These equations
%are equivalent to the Einsten's equations.
The starting point to derive the NC version of the above cosmological
models is the super-Hamiltonian constraint (\ref{3}). 

In order to obtain a NC version for the FRW model, we apply the procedure described in the previous section. Initially, we have to write the zeroth-iterative Lagrangian of the system, which  can be done directly from the super-Hamiltonian in (\ref{3}),
 \begin{equation}\label{eq:4aa}
{\cal L}^{(0)}={P}_a\dot{a}+P_T\dot{T}-V(a,p_a, T,P_T),
 \end{equation}
 where $V=N\Omega$  is the symplectic potential and
\begin{equation}\label{eq:5aa}
\Omega=\frac{-P_a^{2}}{12a}-3ka+\Lambda a^3+P_{T}a^{-3\alpha}.
 \end{equation}
 Notice that the system is treated classically via Symplectic Formalism.  From now on we will follow the steps described in the last section.

In the Lagrangian (\ref{eq:4aa}), the symplectic variables are identified easily as  
$$ \xi ^ {i} = (a, P_a, T, P_T, N)\,\,, $$  
and the corresponding zeroth-iterative one-form  canonical momenta
   \begin{eqnarray}
A_a^{(0)}=P_a&\qquad A_{P_{a}}^{(0)}=0&\qquad A_T^{(0)}=P_T\\\nonumber
A_{P_{T}}^{(0)}=0&\qquad A_N^{(0)}=0&\,\,.
 \end{eqnarray}
Calculating the  one-form canonical momenta using the symplectic matrix definition,
\begin{equation}\label{eq:7a}
f_{\xi^i\xi^j}=\frac{\partial A_{\xi^j}}{\partial{\xi^i}}-\frac{\partial A_{\xi^i}}{\partial{\xi^j}},
\end{equation}
we obtain directly the  zeroth-iteration symplectic matrix, given by
\begin{equation}
f^{(0)}=\left(\begin{array}{ccccc}
0&-1&0&0&0\\
1&0&0&0&0\\
0&0&0&-1&0\\
0&0&1&0&0\\
0&0&0&0&0\end{array}\right).
\end{equation}
However, this matrix is singular, which assumes the existence of constraints in the system, and it has the following zero-mode
\begin{equation}
\nu=\left(\begin{array}{ccccc}
0&0&0&0&1
\end{array}\right).
\end{equation}
Multiplying this zero-mode by the symplectic potential gradient we have that, 
\begin{equation}\label{eq:10a}
\sum_{i=1}^{4}\nu_i\frac{\partial V}{\partial \xi^{i}}=\Omega\,\,,
\end{equation}
where $\Omega$ is a constraint.  This constraint can be introduced
  into the  kinetic  sector of the   zeroth-iterative Lagrangian ${\cal L}^{(0)}$,  through   the Lagrangian multiplier $ \beta $. In this way,  the first-iterative Lagrangian can be written as
\begin{equation}\label{eq:11a}
{\cal L}^{(1)}={P}_a\dot{a}+P_T\dot{T}+\Omega\dot\beta-N\Omega\,\,,
\end{equation}
and the new set of symplectic variables is
$\xi^{i}=(a,P_a,T,P_T,N,\beta)$.   The corresponding first-iterative one-form  canonical momenta $A_{\xi^i}(\xi^j)^{(1)}$  are given by
  \begin{eqnarray}
A_a^{(1)}=P_a&\qquad A_{P_{a}}^{(1)}=0&\qquad A_T^{(1)}=P_T\\\nonumber
A_{P_{T}}^{(1)}=0&\qquad A_N^{(1)}=0&\qquad A_\beta^{(1)}=\Omega.
 \end{eqnarray}

\ni Thus, from the relation (\ref{eq:7a}) we  obtain the
 first-iterative symplectic matrix
\begin{equation}
f^{(1)}=\left(\begin{array}{cccccc}
0&-1&0&0&0&\frac{\partial_\Omega}{\partial a}\\
1&0&0&0&0&\frac{\partial_\Omega}{\partial P_a}\\
1&0&0&0&0&0\\
0&0&1&0&0&\frac{\partial_\Omega}{\partial P_T}\\
0&0&0&0&0&0\\
-\frac{\partial_\Omega}{\partial a}&-\frac{\partial_\Omega}{\partial P_a}&0&-\frac{\partial_\Omega}{\partial P_T}&0&0\end{array}\right).
\end{equation}
However, this matrix is also singular.   If we multiply its zero-mode by the  gradient  of the symplectic potential we will find the same constraint  obtained previously,
\begin{equation}\label{eq:15a}
\sum_{i=1}^{4}\mu_i\frac{\partial
V}{\partial \xi^{i}}=\Omega.
\end{equation}

This result leads us to conclude that the system has a gauge symmetry. In accordance with the symplectic method, this symmetry must be introduced  into the Lagrangian in Eq. (\ref{eq:11a}) through the Lagrange multiplier $\Sigma$.  So, the second-iterative Lagrangian can be written as
\begin{equation}\label{eq:16a}
{\cal L}^{(2)}={P}_a\dot{a}+P_T\dot{T}+\Sigma\dot\eta-N\Omega\,\,,
\end{equation}
where $ \Sigma = N-1 $, i.e., the lapse function is equal to one, which is equivalent to the choice of a physical time, and the new set of symplectic variables is $\xi^{i}=(a,P_a,T,P_T,N,\eta)$.
Using  the symplectic matrix definition in Eq. (\ref{eq:7a}) again,
we have a non-singular second-iterative symplectic matrix. After a straightforward calculation, we obtain  the  inverse of the
 symplectic matrix,
    \begin{equation}\label{eq:18a}
(f^{(2)})^{-1}=\left(\begin{array}{cccccc}
0&1&0&0&0&0\\
-1&0&0&0&0&0\\
0&0&0&1&0&0\\
0&0&-1&0&0&0\\
0&0&0&0&0&-1\\
0&0&0&0&1&0\end{array}\right)\,\,.
\end{equation}
It is important to remember that the elements of this matrix corresponds to the Poisson brackets among the
symplectic variables,  $(f^{-1})^{ij}={\{}\xi^i,\xi^j{\}}$.
The standout moment of the NC introduction through the method described in the last section lies in the assumption that the following relations for the brackets Poisson are,
\begin{eqnarray}\label{eq:19a}
{\{}a,T{\}}&=& \theta\\ \nonumber
{\{}P_{a},P_{T}{\}}&=&\beta.
\end{eqnarray}
These brackets are justified by the fact that $ T $ and its conjugated momentum $P_T  $ can be defined in terms of
specific  entropy and the potential $ \epsilon $ \cite{germano1}, which can be written as a function of energy and the position coordinate. Furthermore, in NC classical mechanics the brackets among the coordinates can be nontrivial. 

With all these values in mind,  using the inverse of the symplectic matrix, including the NC brackets in Eq. (\ref{eq:19a}), we can determine  the symplectic matrix directly
   \begin{equation}\label{eq:abc}
f= \frac{1}{\beta\theta-1}\left [
\begin{array}{cccccc}
0 & 1 & -\beta & 0 & 0 & 0\\
-1 & 0 & 0 & -\theta &0 & 0\\
\beta & 0 & 0 & 1 & 0 & 0\\
0 & \theta & -1 & 0 & 0 & 0\\
0 & 0 & 0 & 0 & 0 & (\beta\theta-1)\\
0 & 0 & 0 & 0 & (1-\beta\theta) & 0\\
\end{array}
\right ] \,\,,
\end{equation}
where $\beta\theta-1\neq 0$, is a constraint. To proceed with the method, we use the symplectic matrix elements (\ref{eq:abc}) and the relations in Eq. (\ref{eq:7a}).  The result is a set of partial differential equations,

\begin{eqnarray}
\frac{\partial A_{{P}_{a}}}{\partial a} - \frac{\partial A_{a}}{\partial P_{a}}&=& \frac{1}{\beta\theta -1} \\\nonumber
\frac{\partial A_{T}}{\partial a} - \frac{\partial A_{a}}{\partial T}&=& \frac{-\beta}{\beta\theta -1} \\\nonumber
\frac{\partial A_{P_{T}}}{\partial P_{a}} - \frac{\partial A_{P_{a}}}{\partial P_{T}}&=& \frac{-\theta}{\beta\theta -1} \\\nonumber
\frac{\partial A_{{P}_{T}}}{\partial T} - \frac{\partial A_{T}}{\partial P_{T}}&=& \frac{1}{\beta\theta -1} \\\nonumber
\frac{\partial A_{{\eta}}}{\partial N} - \frac{\partial A_{N}}{\partial \eta}&=& 1 .\\\nonumber
\end{eqnarray}

\ni The system above have one possible and convenient solution,

\begin{eqnarray}
A_{a}&=&\frac{1}{1-\beta\theta}(P_{a}-\beta T), \qquad  A_{P_{a}}=\frac{\theta}{\beta\theta-1}P_{T}\\\nonumber
A_{T}&=&\frac{1}{1-\beta\theta}P_{T}, \qquad \qquad \qquad A_{P_{T}}=0 \\\nonumber
A_{\eta} &=& \Sigma    \qquad \qquad    \qquad \qquad                     \qquad A_N=0 \nonumber .
\end{eqnarray}

\ni Therefore,  from the  one-form canonical momenta above the new NC first-order Lagrangian  can be computed directly. However, we also have to consider  that the model remains second-order in velocities.  
%$$A_{P_{T}}=0 \qquad \mbox{and} \qquad A_{P_{a}}=\frac{\theta}{\beta\theta-1}P_{T}=0,$$then $ \theta = 0 $. 
Consequently, we will use the canonical momenta given by
\begin{eqnarray}
A_{a}&=&\frac{1}{1-\beta\theta}(P_{a}-\beta T)  \\\nonumber
A_{T}&=&\frac{1}{1-\beta\theta}P_{T}          \\\nonumber
A_{\eta} &=& \Sigma \,\,.                             
\end{eqnarray}
The respective NC first-order Lagrangian is,
\begin{equation}\label{eq:27a}
{\cal L}= \frac{1}{1-\beta\theta}(P_{a}-\beta T)\dot{a}  +\frac{1}{1-\beta\theta}P_{T}\dot{T}+\Sigma \dot{\eta}-V(\xi)\,\,,
\end{equation}
where the symplectic potential is written as
\begin{equation}
V(\xi)=\frac{-P_a^{2}}{12a}-3ka+\Lambda a^3+P_{T}a^{-3\alpha}.
\end{equation}

 In order to obtain a commutative first-order   Lagrangian, we propose a coordinate transformation in  the classical phase space, analogous to the shift-operator 
$\hat{x}_i=X_i+\frac{1}{2}\theta_{ij}\hat{p}^j$  used in NC Quantum Mechanics (NCQM), given by
\begin{eqnarray}\label{eq:29}
\widetilde{P}_{a}=\frac{P_{a}-\beta T}{1-\beta\theta},\qquad & \widetilde{P}_{T}=\frac{P_{T}}{1-\beta\theta}.\\\nonumber
\end{eqnarray}
Applying the  transformation above in (\ref{eq:27a}), we obtain the new commutative first-order Lagrangian,
  \begin{equation}\label{eq:90}
 \widetilde{{\cal L}}=\widetilde{P}_{a}\dot{a}  + \widetilde{P}_{T}\dot{T}+\Sigma \dot{\eta}-V(\xi)\,\,,
 \end{equation}
 where the symplectic potential 
   \begin{equation}
V(\xi)=\frac{-P_a^{2}}{12a}-3ka+\Lambda a^3+P_{T}a^{-3\alpha}
\end{equation}
is written in NC coordinates that satisfy the usual Poisson brackets given in Eq. (\ref{eq:19a}).
It is important to note that despite the variables in the Lagrangian Eq. (\ref{eq:90}) are commutative, there are NC contributions  into the symplectic potential.  Finally,   the modified super-Hamiltonian of the  system    is identified as being the symplectic potential, so it can be written as
 \begin{equation}\label{eq:hnc}
H=-\frac{(\widetilde{P}_a+\beta T)^{2}}{12a}-3ka+\Lambda a^3+\widetilde{P}_{T}a^{-3\alpha}.
 \end{equation}

\ni Notice that when $\theta=\beta=0$ from (\ref{eq:19a}) and (\ref{eq:29}), we recover, from (\ref{eq:hnc}), the Hamiltonian in (\ref{3}).

\section{Classical behavior of the NC cosmological models}
\label{section4}

In order to investigate the contributions coming from the noncommutativity between
the canonical variables and momenta in the classical FRW cosmological models, 
we derive the dynamical equations by computing the Hamilton's equations for the
total Hamiltonian $NH$, where $N$ is the lapse function and $H$ is the modified 
super-Hamiltonian Eq. (\ref{eq:hnc}). We also use the constraint equation $H = 0$. 
The new momenta $\widetilde{P}_a$ and $\widetilde{P}_{T}$, present in the $H$, are 
given by Eq. (\ref{eq:29}). In those expressions of $\widetilde{P}_a$ and $\widetilde{P}_{T}$,
$\theta$ and $\beta$ are the parameters associated with the noncommutativity 
among the canonical variables and momenta, respectively (Eq. (\ref{eq:19a})). In the 
present study, we are going to consider only the contribution coming from the parameter 
$\beta$. In other words, we shall fix $\theta = 0$. The reason is because we do not want 
the resulting dynamical equations depending on terms of order greater than two in the 
velocities.

Assuming that $N=1$, we obtain the following two dynamical equations for the scalar factor $a$,

\begin{equation}
\label{eq:EFM1}
\frac{\dot{a}^2}{a^2}+\frac{k}{a^2}=\frac{\Lambda}{3}+\frac{\rho}{3}-
\frac{\beta}{3} a^{-3\alpha-2},
\end{equation}
\begin{equation}
\label{eq:EFM2}
\frac{2\ddot{a}}{a}+\frac{\dot{a}^{2}}{a^{2}}+\frac{k}{a^{2}}=\Lambda - p -
\beta a^{-3\alpha-2}(\frac{1}{3}-\alpha),
\end{equation}
where the dot means derivative with respect to the time coordinate $t$, in the present
gauge. The first equation is the generalization for the NC models of the
Friedmann equation. Both equations reduce the corresponding dynamical equations of the
commutative models when we use that $\beta = 0$.

It will be very useful to rewrite the generalized Friedmann equation (\ref{eq:EFM1})
 with the following form,

\begin{equation}
\label{eq:efm1}
{\dot{a}^2}+V(a)=0,
\end{equation}
where
\begin{equation}
\label{eq:c3}
V(a)=k-\frac{1}{3}\Lambda{a^2}+\frac{\beta}{3}a^{-3\alpha}-\frac{C}{3}a^{-3\alpha-1},
\end{equation}

\ni where $C$ is a positive integration constant which is related to the initial fluid energy
density ($\rho_0$).   We notice that the total energy of this conservative system is equal
to zero. From the observation of the potential curve $V(a)$, we shall be able to derive the
qualitative dynamical behavior of $a(t)$.

It is important to notice that the NC models described by the two dynamical
equations above satisfy the energy conservation equation. In order to show this result, we use both Eqs. (\ref{eq:EFM1}) and (\ref{eq:EFM2}) to obtain,

\begin{equation}
\dot{\rho}+3\frac{\dot{a}}{a}(\rho+P)=0.
\end{equation}
This is the energy conservation equation to the commutative version of the models. Therefore,
the noncommutativity does not violate the energy conservation law.

Both equations (\ref{eq:EFM1}) and (\ref{eq:EFM2}) are not independent. 
%All solutions $a(t)$ to Eq. (\ref{eq:EFM1}) are also solutions to Eq. (\ref{eq:EFM2}). On the other hand, not all solutions to Eq. (\ref{eq:EFM2}) are solutions to Eq. (\ref{eq:EFM1}). But all solutions to Eq. (\ref{eq:EFM2}) are solutions to,
On the other hand, it can be shown that the set of solutions of Eq. (\ref{eq:EFM2}) is not the same set of solutions of Eq. (\ref{eq:EFM1}).   However, the set of solutions of Eq. (\ref{eq:EFM2}) is the same set of solutions of

\begin{equation}
\label{eq:5000}
\frac{\dot{a}^2}{a^{2}}+\frac{k}{a^2}=\frac{8\pi G\rho}{3}+\frac{\Lambda}{3}-
\frac{\beta}{3} a^{-3\alpha-2}+f(t),
\end{equation}

\ni where $f(t)=f_{0}a^{-3}(t)$ and $f_0$ is an integration constant. We may impose that the solutions of Eq. (\ref{eq:EFM2}) be, also, solutions of Eq. (\ref{eq:EFM1}). It can be accomplished by
fixing the following initial condition on the velocity $\dot{a}(t=t_0)\equiv\dot a_0$,

\begin{equation}
\label{eq:ic}
\dot{a}_0=\mp\bigg{\{}\bigg{(}-\frac{k}{a^2}+\frac{\Lambda}{3}+\frac{8\pi G\rho}{3}-\frac{\beta}{3} a_0^{-3\alpha-2}\bigg{)}a_0^{2}\bigg{\}}^\frac{1}{2},
\end{equation}

\ni where $a(t=t_0)\equiv a_0$. The only way this initial condition satisfy Eq. (\ref{eq:5000})
is when $f_0 = 0$. Since this result must be valid for all times this initial condition guarantees that all solutions $a(t)$ of Eq. (\ref{eq:EFM2}) are also solutions of Eq. (\ref{eq:EFM1}). In the following analysis we shall restrict our attention to the positive sign in Eq. (\ref{eq:ic}).

In order to derive the scalar factor as a function of $t$, we shall, initially, observe the potential
curve $V(a)$ from Eq. (\ref{eq:EFM1}) and then solve Eq. (\ref{eq:EFM2}). Unfortunately, for generic
values of the different parameters present in Eq. (\ref{eq:EFM2}), this equation does not have algebraic
solutions. Therefore, we shall solve it numerically using a Fehlberg
fourth-fifth order Runge-Kutta method with degree four interpolant, for each different values of the
parameters. 

We have four parameters in Eq. (\ref{eq:EFM2}). The first one is $k$, which is associated with
the curvature of the spatial sections and we may assume three different values: -1, 0, +1. The second one is $\Lambda$, the cosmological constant, which can be positive, negative or zero. The third one is
$\beta$, the NC parameter, which can be also positive, negative or zero. It is important to mention
that this last case $\beta = 0$ means that the model is commutative. 
Finally, we have the parameter $\alpha$,
which is present in the equation of state for the perfect fluid ($p = \alpha \rho$). Each value
of $\alpha$ defines a different perfect fluid. Here, we shall consider six different values of $\alpha =
(1, 1/3, 0, -1/3, -2/3, -1)$, which represents respectively: stiff matter, radiation, dust, cosmic strings,
domain walls and dark energy. Taking into account all possible values of the parameters, we have considered
162 qualitatively different cases and we solved Eq. (\ref{eq:EFM2}) for all of them. In fact, we have solved
Eq. (\ref{eq:EFM2}) for a number of cases greater than 162 because we have considered not only the different signs for the cases of $\Lambda$ and $\beta$ but also for different absolute values of these parameters.

After solving Eq. (\ref{eq:EFM2}) for all possible cases, mentioned above, we have reached the following
general conclusions. If the parameter $\beta$ is positive, it has the net effect of producing a negative
acceleration in the scalar factor evolution. In cases where the universe is expanding, the presence of
a positive $\beta$, will slow the expansion or even stops it and forces the universe to contract. If the
parameter $\beta$ is negative, it has the net effect of producing a positive acceleration in the scalar
factor evolution. In cases where the universe is expanding, the presence of a negative $\beta$, will
increase the expansion speed. On the other hand, in cases where the Universe is contracting, the presence
of a negative $\beta$ may force the Universe to expand. From these observations, we notice that $\beta$
modifies the acceleration of the scale factor in the opposite way that the cosmological constant does.
Next, we shall present some particular cases where the above conclusions can be clearly verified.

\subsection{Universe with $k = 0$, $\alpha = 0$ and $\Lambda > 0$}

For this case the Universe has flat spatial sections, it is filled with dust and the cosmological
constant is positive. The potential $V(a)$ in Eq. (\ref{eq:c3}) is given by,

\begin{equation}
\label{eq:fmb}
V(a)=-\frac{\Lambda a^{2}}{3}-\frac{C}{3a}+\frac{\beta}{3}.
\end{equation}

From the potential expression it is easy to see that this universe starts to expand from a
singularity at $a = 0$. For,
\begin{equation}
\label{condition}
\beta < (27C^2\Lambda/4)^{(1/3)},
\end{equation}
the Universe expands initially in a decelerated rate and later on in an accelerated rate toward
$a \to \infty$. Since, in the present case $C, \Lambda > 0$, the above condition is satisfied for
all cases where $\beta \leq 0$. The condition Eq. (\ref{condition}) guarantees that the potential
$V(a)$ Eq. (\ref{eq:fmb}) is always negative. For $\beta = 0$, we have the commutative solution. If
$\beta$ is positive and satisfies Eq. (\ref{condition}), the universe expands in a smaller rate than
in the commutative case. On the other hand, if $\beta$ is negative the universe expands in a greater rate
than in the commutative case. As an example we show in figure 1 the potential $V(a)$, Eq. (\ref{eq:fmb}).
For the case where $\Lambda = 0.01$, $C =0.01$ and $\beta = 0.01(red), 0(blue), -0.01(green)$. Then, in figure 2, we show the solutions to Eq. (\ref{eq:EFM2}) for the present case with the same values of the
parameters of figure 1.

\begin{figure}[H]
\begin{center}
\includegraphics[width=2.31in,height=2.22in]{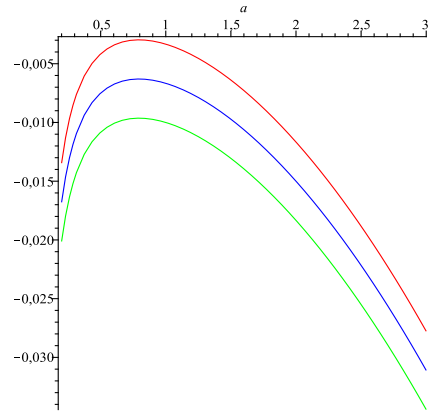}
\caption{$V(a)$ Eq. (\ref{eq:fmb}), for $C = 0.01$, $\Lambda = 0.01$ and
$\beta=0.01(red), 0(blue), -0.01(green)$.}
\end{center}
\label{fig1}
\end{figure}

\begin{figure}[H]
\begin{center}
\includegraphics[width=2.31in,height=2.22in]{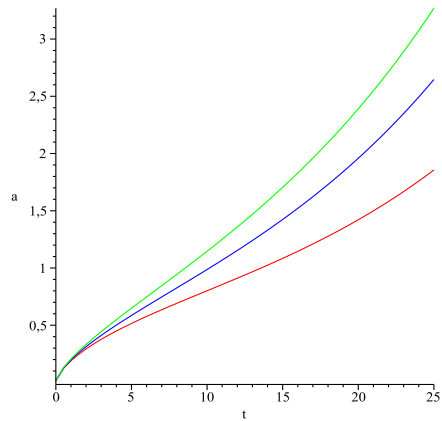}
\caption{Solutions $a(t)$ to Eq. (\ref{eq:EFM2}), for $ C = 0.01$, $k = 0$, $ \Lambda = 0.01$,
$\alpha = 0$ and $\beta = 0.01(red), 0(blue), -0.01(green)$.}
\end{center}
\label{fig2}
\end{figure}

\clearpage

\subsection{Universe with $k = 0$, $\alpha = 0$ and $\Lambda < 0$}

For this case the universe has flat spatial sections.   It is filled with dust and the cosmological constant is negative. The potential $V(a)$ Eq. (\ref{eq:c3}) is given by,

\begin{equation}
\label{eq:fmb1}
V(a)=\frac{|\Lambda| a^{2}}{3}-\frac{C}{3a}+\frac{\beta}{3},
\end{equation}
where $|\Lambda|$ is the cosmological constant absolute value.

From the expression for the potential (\ref{eq:fmb1}) it is easy to see that this universe starts to expand from a singularity at
$a = 0$. Then, for all possible values of $\beta$ the scale factor reaches a maximum value and
starts to contract toward a {\it big crunch} singularity at $a = 0$. For $\beta = 0$, we have
the commutative solution. If $\beta$ is positive the scale factor is forced to contract
in a stronger way than in the commutative case. Here, the maximum value of $a$ is smaller
than in the commutative case. On the other hand, if $\beta$ is negative the scale factor will
contract in a weaker way than in the commutative case. Here, the maximum value of $a$ is
greater than in the commutative case. As an example we show in figure 3 the potential $V(a)$,
Eq. (\ref{eq:fmb1}), for the case where $\Lambda = - 0.01$, $C = 0.01$ and $\beta = 0.01(red), 0(blue),
-0.01(green)$. Then, in figure 4, we show the solutions to Eq. (\ref{eq:EFM2}) for the
present case with the same values of the parameters of figure 3.

\begin{figure}[H]
\begin{center}
\includegraphics[width=2.31in,height=2.22in]{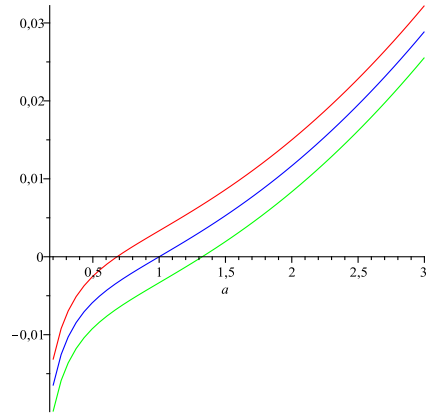}
\label{fig3}
\caption{$V(a)$ Eq. (\ref{eq:fmb1}), for $C = 0.01$, $\Lambda = - 0.01$ and
$\beta = 0.01(red), 0(blue), -0.01(green)$.}
\end{center}
\end{figure}

\begin{figure}[H]
\begin{center}
\includegraphics[width=2.31in,height=2.22in]{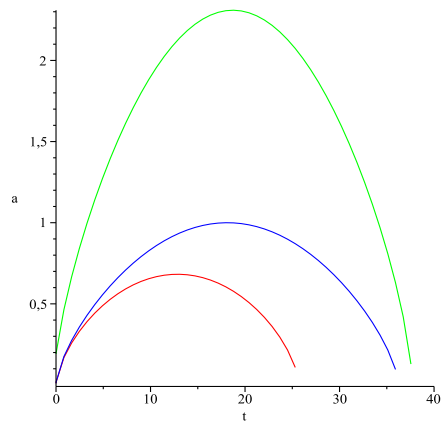}
\label{fig4}
\caption{Solutions $a(t)$ to Eq. (\ref{eq:EFM2}), for $C = 0.01$, $k = 0$, $ \Lambda = - 0.01$,
$\alpha = 0$ and $\beta = 0.01(red), 0(blue), -0.01(green)$.}
\end{center}
\end{figure}

\clearpage

\subsection{Universe with $k = 0$, $\alpha = 1/3$ and $\Lambda = 0$}

For this case the Universe has flat spatial sections, it is filled with radiation and there is no
cosmological constant. The potential $V(a)$ Eq. (\ref{eq:c3}) is given by,

\begin{equation}
\label{eq:fmb2}
V(a)=-\frac{C}{3a^2}+\frac{\beta}{3a}.
\end{equation}

Here, we have a very interesting situation. For $\beta \leq 0$  the universe starts to expand
from a singularity at $a = 0$. It expands in a decelerated rate until it reaches asymptotically
$a \to \infty$. For $\beta = 0$, we have the commutative solution. This commutative case is well-know
in the literature and it corresponds to Friedmann equation Eq.(\ref{eq:efm1}) that can be solved to
give the algebraic solution: $a(t) = At^{(1/2)}$, where $A$ is a constant. If $\beta$ is negative
the universe expands in a rate greater than the commutative case. On the other hand,
if $\beta$ is positive the universe expands up to a maximum size given by $c/\beta$, then it is
forced to contract toward a {\it big crunch} singularity at $a = 0$. Therefore, we see that the
NC parameter may change drastically the universe evolution. As an example we show in
figure 5 the potential $V(a)$, Eq. (\ref{eq:fmb2}), for the case where $\Lambda = - 0.01$,
$C = 0.01$ and $\beta = 0.01(red), 0(blue), -0.01(green)$. Then, in figure 6, we show the
solutions to Eq. (\ref{eq:EFM2}) for the present case with the same values of the parameters of
figure 5. It is important to mention that the term depending on $\beta$ in Eq.
(\ref{eq:EFM2}) is not present in this case. The solutions of Eq. (\ref{eq:EFM2}) depend on
$\beta$, in the present case, due to the initial condition Eq. (\ref{eq:ic}).

\begin{figure}[H]
\begin{center}
\includegraphics[width=2.31in,height=2.22in]{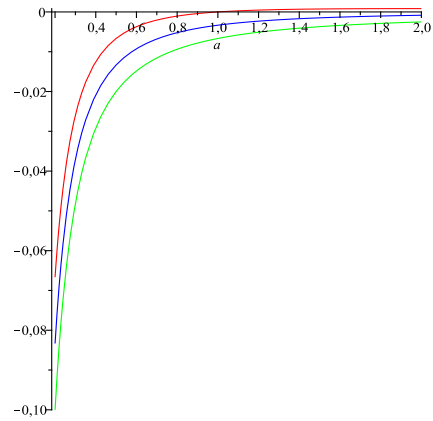}
\label{fig5}
\caption{$V(a)$ Eq. (\ref{eq:fmb2}), for $ C = 0.01$ and
$\beta = 0.01(red), 0(blue), -0.01(green)$.}
\end{center}
\end{figure}

\begin{figure}[H]
\begin{center}
\includegraphics[width=2.31in,height=2.22in]{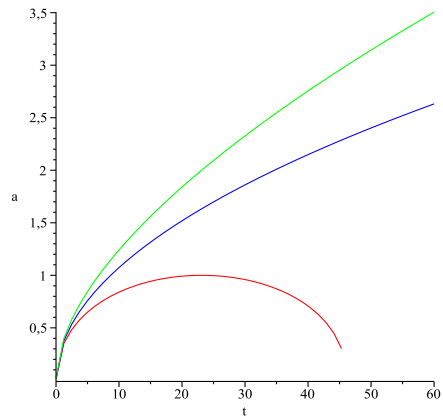}
\label{fig6}
\caption{Solutions $a(t)$ to Eq. (\ref{eq:EFM2}), for $C = 0.01$, $k = 0$, $\Lambda = 0$,
$\alpha = 1/3$ and $\beta = 0.01(red), 0(blue), -0.01(green)$.}
\end{center}
\end{figure}

\clearpage

\subsection{Universe with $k = -1$, $\alpha = -2/3$ and $\Lambda < 0$}

For this case the Universe has constant negatively curved spatial sections, it is filled
with a domain wall perfect fluid and there is a negative cosmological constant. The potential
$V(a)$ in Eq. (\ref{eq:c3}) is given by,

\begin{equation}
\label{eq:h1h}
V(a)=-1-\frac{\Lambda}{3}a^{2}+\frac{\beta}{3}a^{2}-\frac{C}{3}a.
\end{equation}

Here, we have the situation opposite to the previous case. For $\beta > \Lambda$, the universe
starts to expand from $a = 0$, which is not a singularity. Then, the scale factor reaches a maximum
value, $$a_{max} = (C+\sqrt{C^2+12(\beta - \Lambda)})/[2(\beta - \Lambda)]\,\,,$$ and starts to contract
toward $a = 0$. 

Since $\Lambda < 0$, the Universe is bounded for the commutative solution. If
$\beta$ is positive the scale factor is forced to contract in a stronger way than in the commutative
case. Here, the maximum value of $a$ is smaller than in the commutative case. 

On the other hand, if
$\beta \leq \Lambda$ the Universe starts to expand from $a = 0$ and accelerates its expansion toward
$a \to \infty$. Therefore, we will see another example where the NC parameter may change
drastically the universe evolution. As an example we show in figure 7 the potential $V(a)$,
Eq. (\ref{eq:h1h}), for the case where $\Lambda = - 0.01$, $C = 0.01$ and
$\beta = 0.011(red), 0(blue), -0.011(green)$. Here, we can find an algebraic solution for the scalar
factor as a function of time. Using Eqs. (\ref{eq:EFM1}) and (\ref{eq:EFM2}), and using the values of
the parameters for the present case, we obtain the following differential equation,

\begin{equation}
\label{eq:efm5}
{\ddot{a}} + \omega^2\,a = \frac{C}{6}.
\end{equation}
%{\ddot{a}} + \bigg{(}\frac{\beta}{3} - \frac{\Lambda}{3}\bigg{)}a = \frac{C}{6}.

The general solution of this equation is given by,

\begin{equation}
\label{eq:solu}
a(t)= Ae^{i \omega t} + Be^{-i\omega t} + \frac{C}{6{\omega}^{2}}.
\end{equation}
where $A$, $B$ are integration constants and $\omega^2 = (\beta - \Lambda)/3$.

If $\omega^2 > 0$ Eq. (\ref{eq:efm5}) represents a driven harmonic oscillator with a constant driven
force. The solution, Eq. (\ref{eq:solu}), is oscillatory. On the other hand, if $\omega^2 < 0$ Eq.
(\ref{eq:efm5}) represents an unbounded system and the solution Eq. (\ref{eq:solu}) grows exponentially.

\begin{figure}[H]
\begin{center}
\includegraphics[width=2.31in,height=2.22in]{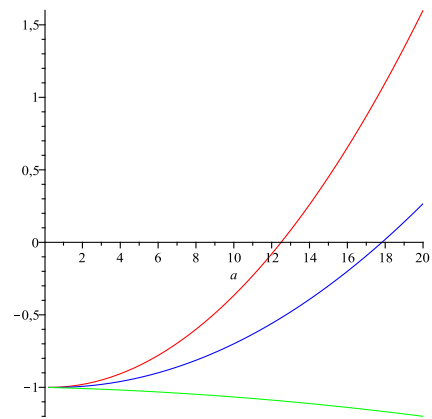}
\label{fig7}
\caption{$V(a)$ Eq. (\ref{eq:h1h}), for $\Lambda = - 0.01$, $C = 0.01$ and
$\beta = 0.01(red), 0(blue), -0.01(green)$.}
\end{center}\end{figure}

\clearpage

\section{The noncommutativity and the Universe evolution}
\label{section5}

Taking into account the above results we can see that the presence of $\beta$ in the cosmological equations
may modify in a fundamental way the evolution of the Universe. In order to gain a better insight about the role
of $\beta$, let us rewrite the generalized Friedmann equation (\ref{eq:EFM1}) in the following way,

\begin{equation}
\label{eq:hh}
\frac{\dot{a}^2}{a^2}+\frac{k}{a^2}=\frac{8\pi G\rho}{3}+\widetilde{\Lambda} (a),
\end{equation}
where $$\widetilde{\Lambda} (a)= (1/3)(\Lambda - \beta a^{-3\alpha-2})\,\,.$$
We may interpret $\widetilde{\Lambda}$ as a function of the scale factor that generalizes the
cosmological constant. It is important to notice that, the contributions coming from $\Lambda$ and
the term depending on $\beta$ oppose to each other. In order for them to reinforce each other, $\Lambda$ and
$\beta$ must have opposite signs.

Depending on the value of $\alpha$, we may have two entirely different scenarios for the evolution of the
Universe described by $\widetilde{\Lambda}$.

\subsection{First scenario}

For $\alpha > - 2/3$, the modulus of the term $\beta a^{-3\alpha-2}$
decreases when $a$ increases. This behavior agrees with the idea that noncommutativity should be most
important at the beginning of the Universe. After some time, when the universe has evolved considerably
$a$ is large enough so that $\widetilde{\Lambda} \approx \Lambda/3$. From this moment the Universe
evolves without any remembrance of a NC phase.

\subsection{Second scenario}

For $\alpha < - 2/3$, the modulus of the term $\beta a^{-3\alpha-2}$
increases when $a$ increases. After some time, when the Universe has evolved considerably $a$ is large
enough so that $\widetilde{\Lambda} \approx -\beta a^{-3\alpha-2}/3$. From this moment the Universe evolution is dominated by the NC term. 

If $\beta > 0$, the Universe will eventually reach a
maximum size and contract again toward $a \to 0$. On the other hand, if $\beta < 0$, the Universe will
expand forever in an accelerated rate. Here, we have a very interesting possibility. It is possible to
consider the NC effects as a candidate to explain the present accelerated expansion of the
Universe \cite{riess},\cite{perlmutter}. In fact, some authors have already considered this possibility
using classical, NC cosmological models different from ours 
\cite{pedram},\cite{obregon},\cite{neves}.

As a matter of completeness, we observe that for $\alpha = - 2/3$, $\widetilde{\Lambda}$ is a constant.
It is given by, $\widetilde{\Lambda} = (1/3)(\Lambda - \beta)$. In this case, the NC parameter
can be compared with the cosmological constant. If $\Lambda = 0$, the NC parameter plays
the role of a cosmological constant with opposite sign.

\clearpage

\section{Final considerations and conclusions}
\label{section6}

One of the mysteries that defies the theoretical physics today is how to promote the unification of two independent sets of very different concepts.  Namely, one set that rules the microscopic world, i.e., the quantum mechanics and the other that govern the macroscopic Universe, i.e., general relativity.  The main motivation to unify both disconnected (so far) sets is to understand the physics of the early Universe and the physics that rules the origin and the physical structure of the black holes.  In few words, we are looking for the generalized idea of the so called quantum gravity.

As we know the NC parameter has its value measured at the Planck scale.  Besides, the string theory, a candidate to carry out this unification, embedded in a magnetic background (to explain it in a nutshell), was found to have a NC algebra.  In view of these two theoretical facts it is natural to believe that to investigate NC theories can be one of the adequate paths to conduct us to such unification.

It is the main objective of this work to pursue this target.  To accomplish this, we used the generalized NC symplectic formalism to introduce naturally noncommutativity inside the equations that provide the dynamics of the Universe, i.e., the Friedmann equations.  With this procedure, we introduced a Planckian object inside the classical equations of motion of the Universe.

We have constructed NC versions of several Friedmann-Robertson-Walker cosmological models.
These models may have positive, negative or zero spatial sections curvatures.  
They are coupled to different types of perfect fluids and they may have a positive, negative or zero cosmological constant. We have used
the Schutz variational formalism in order to write the proper Hamiltonian for the models. We have performed a complete analysis of the evolution of the Universes described by each model. We solved the dynamical equations for the scale factor. 

Based on the results obtained here we have concluded that, if the parameter $\beta$ is positive, it has the final
effect of producing a negative acceleration in the scalar factor evolution. In cases where the Universe is
expanding, the presence of a positive $\beta$, will slow the expansion or even stops it and forces the
Universe to contract. If the parameter $\beta$ is negative, it has the effect of producing a positive
acceleration in the scalar factor evolution. In cases where the Universe is expanding, the presence of a
negative $\beta$, will increase the expansion speed. On the other hand, in cases where the Universe is
contracting, the presence of a negative $\beta$ may force the Universe to expand. From these observations,
we notice that $\beta$ modifies the acceleration of the scale factor in the opposite way that the cosmological
constant does. We have also presented some particular cases where the above conclusions could be clearly
verified. Finally, based on our results, we construct a function ($\widetilde{\Lambda}$) that depends on $a$,
$\beta$ and $\alpha$ and generalizes the cosmological constant. We have discussed the possible scenarios for
the Universe evolution predicted by $\widetilde{\Lambda}$, depending on the value of $\alpha$.

\acknowledgments 

EMCA would like to thank the kindness of UFJF's Physics Department.
G. Oliveira-Neto thanks FAPEMIG, CNPq and FAPERJ for partial financial support.

\end{document}